\documentclass[preprint,preprintnumbers,amsmath,amssymb]{revtex4}

\usepackage{graphicx}
\usepackage{dcolumn}
\usepackage{bm}

\begin{document}

\title{Analysis of fractional Gaussian noises using level crossing method}
\author{M. Vahabi, G. R. Jafari, M. Sadegh Movahed\\
\emph{Department of Physics, Shahid Beheshti University, G. C.,
Evin, Tehran 19839, Iran}}

\begin{abstract}

The so-called level crossing analysis has been used to investigate
the empirical data set. But there is a lack of interpretation for
what is reflected by the level crossing results. The fractional
Gaussian noise as a well-defined stochastic series could be a
suitable benchmark to make the level crossing findings more sense.
In this article, we calculated the average frequency of upcrossing
for a wide range of fractional Gaussian noises from logarithmic
(zero Hurst exponent, $H=0$), to Gaussian, $H=1$, ($0<H<1$). By
introducing the relative change of the total numbers of upcrossings
for original data with respect to so-called shuffled one,
$\mathcal{R}$, an empirical function for the Hurst exponent versus
$\mathcal{R}$ has been established. Finally to make the concept more
obvious, we applied this approach to some financial series.

\end{abstract}
\maketitle

\section{Introduction}

We are living in a world in which processes of stochastic type are
ubiquitous. Although the random values of a stochastic process at different
times may be independent random variables, in most commonly
considered situations they exhibit complicated statistical
correlations. Thus, over past decades several different methods have
been introduced to investigate the properties of these processes.
To the best of out knowledge all of these methods need a scaling
relation in order to measure the information included in the series.
Another well-known method is the level crossing method (LC) in which
no scaling feature is explicitly required
\cite{rice44,newland,Tabar1,Jafari,tabar2,jensen} and this is the main
advantage of this method in estimating the statistical information
of the series.

What is the reason that level crossing method has been created? This
method was invented to study the series with different insight. By
level crossing, we can measure the memory, non-Gaussianity and
waiting time (length) (an average time (length) interval that we
should wait for an event to take place again
\cite{level,Johansen,Bunde,Newell,mahsa}). In order to find the way
that correlations enters the level crossing method, we have
considered some fractional Gaussian noises (fGns) (generalization of
ordinary discrete white Gaussian noise) whose Hurst exponents and
distribution functions are known. Indeed, the so-called Hurst
exponent, $H$, gives a quantitative measure of the long-term
persistence of a signal. In particular, the exponents $0<H<0.5$ and
$0.5<H<1$ correspond to negative (anti-correlation) and positive
correlation, respectively, while $H=0.5$ represents an uncorrelated
Gaussian process. Since fGns are well known examples, their
comparison with empirical data can be used as a criterion to better
understand the results obtained from the level crossing method
applied to unknown empirical data.

Here, an integrated quantity namely, $N^+_{tot}$ which represents
the total number of upcrossings of a typical series reflects how
memory plays role. For a better investigation of the memory effects,
we have calculated shuffled counterparts of each of underlaying time
series and compared their associated total number of
 so-called crossings, ($N^+_{sh}$), with that of given by their original time
series, $N^+_{tot}$, to obtain the percentage of the change in the
system. By the shuffling procedure, autocorrelations are destroyed.
Finally, we have considered the data from some markets and mentioned
technique has been  applied to daily log-returns of three time
series of financial markets named $S\&P 500$, Dow Jones and Tehran
Stock Exchange (TSE) in the same time interval from $4$ Jan $2006$
to $4$ Jan $2010$ \cite{data}. Although it is said that $S\&P 500$
and Dow Jones have Hurst exponent of about $0.5$, the results show
that they are not exactly white noises. Actually memory exists which
have not been observed. Also, the source for this inconsistency can
be obtained (table \ref{Tb1}).

This paper is organized as follows. In Section II, we will give a
brief history and explanations concerning the level crossing method.
Then, data description and analysis based on this method for
different fractional Gaussian noises and also its application to
some empirical data are given in Section III. In Section IV, we
present our conclusion.

\begin{figure}[t]
\includegraphics[width=11cm,height=9cm,angle=0]{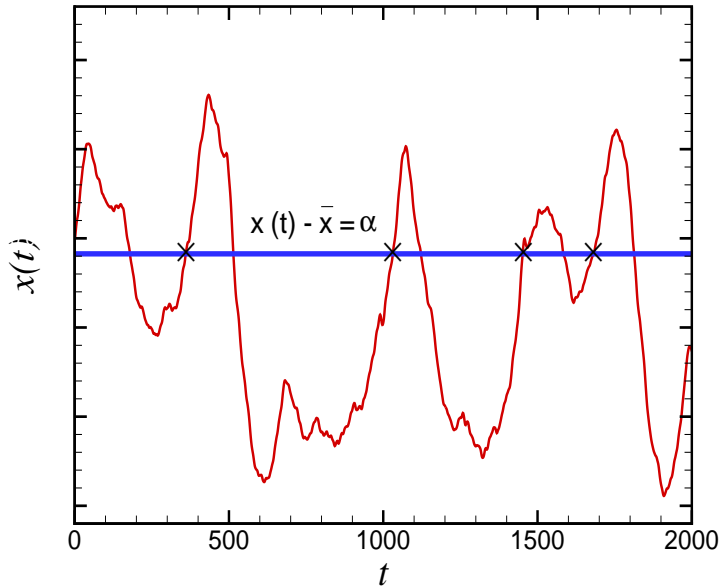}
\caption{Schematic of upcrossing for an arbitrary level, $x(t)-\bar{x}=\alpha$.}\label{fig1}
\end{figure}

\section{Level Crossing Analysis}
\begin{figure}[t]
\includegraphics[width=11cm,height=9cm,angle=0]{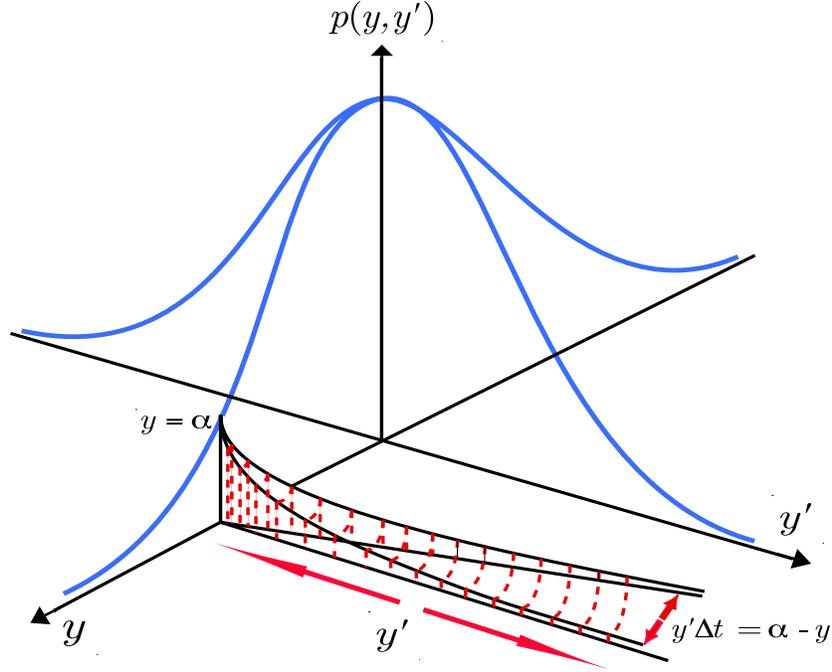}
\caption{A schematic of joint probability density function of
underlaying series and its derivative with respect to dynamical
parameter that here is time. The shaded area corresponds to desired
region satisfying in Eq. (3) \cite{newland,sadegh11}. }
\label{joint}
\end{figure}

For the sake of clarity, we begin with a summary of the LC analysis
\cite{rice44,newland,Tabar1,Jafari,tabar2,jensen}. Consider a
typical time series $\{x(t)\}$ and $n_{\alpha}^{+}$ denotes the
number of positive difference crossings (upcrossings) at level
$x(t)- \bar x = \alpha$ in time interval $T$ (see Fig. \ref{fig1}).
The mean value of $n_{\alpha}^{+}$ for all time intervals be
$N_{\alpha}^{+}(T)$ \cite{Tabar1},
\begin{equation}
N_{\alpha}^{+}(T)=\langle n_{\alpha}^{+}(T)\rangle,
\end{equation}
here $\langle . \rangle$, represents ensemble average. For a
homogeneous (stationary) process, the average number of upcrossings
is proportional to the time interval $T$. Subsequently,
\begin{equation}
N_{\alpha}^{+}(T)=\nu^{+}_{\alpha} T,
\end{equation}
in which $\nu_{\alpha}^{+}$ is the average frequency of upcrossings
at the level equates to $y = x(t)- \bar x = \alpha$. The frequency
parameter $\nu_{\alpha}^{+}$ could be deduced from the underlying
joint probability distributions of $y \equiv x(t)- \bar x$ and
$y'=\frac{y(t+\Delta t)-y(t)}{\Delta t}=\frac{\Delta y}{\Delta t}$,
namely $p(y,y') $ \cite{rice44,newland,Tabar1,Jafari,tabar2}. To
this end, consider a time scale $\Delta t$ of a typical sample
function, if $x(t)-\bar x < \alpha$ at time $t$ and $x(t)- \bar x >
\alpha$ at $t+\Delta t$ or alternatively the changes in $x(t)$ is
positive in the time interval $\Delta t$, there will be a positive
crossing of $x(t)-\bar x =\alpha$. In an other word, two following
necessary and sufficient conditions should be satisfied to have an
upcrossing in time interval, $\Delta t$

\begin{eqnarray} x(t)- \bar x &<& \alpha \hspace{.6cm}\\ \nonumber
 \hspace {.6cm} \frac{\Delta\left[x(t)-\bar x\right]}{\Delta t}&>&
\frac{\alpha-\left[x(t) - \bar x\right] }{\Delta t}.
\end{eqnarray}\label{condition}

\begin{figure}[t]
\includegraphics[width=11cm,height=9cm,angle=0]{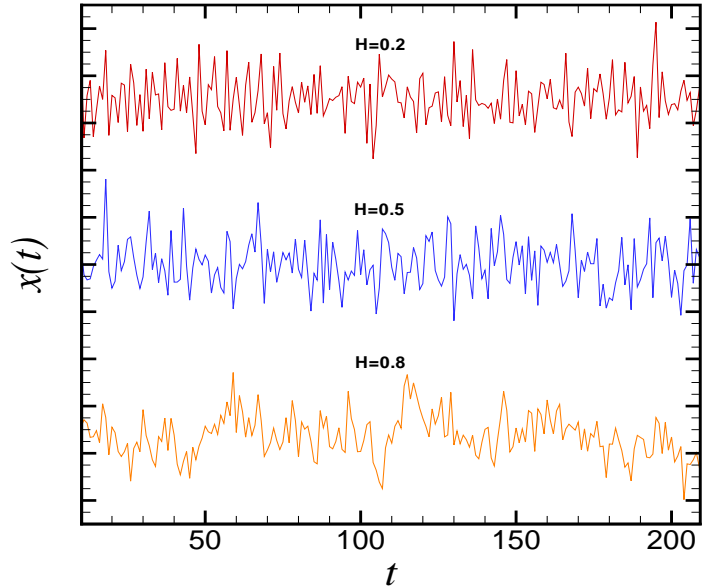}
\caption{Time series of three fractional Gaussian noises with Hurst
exponents $H$, $0.2, 0.5$ and $0.8$ from up to down.}\label{fig2}
\end{figure}

In order to examine whether  the above conditions are satisfied at
any arbitrary time $t$, we should find how the values of $y\equiv
x(t)- \bar x$ and corresponding time derivative are distributed by
considering their joint probability density $p(y,{y}^{\prime})$.
Suppose that the level $y=\alpha$ and interval $\Delta t$ are
specified. Then, we are only interested in values of $y < \alpha$
and values of ${y}^{\prime}
> \frac{\alpha-y}{\Delta t}$, corresponding the region between
the lines $y\leq \alpha$ and ${y}^{\prime}\geq
\frac{\alpha-y}{\Delta t}$ in the plane of ($y,{y}^{\prime}$) (see
Fig. \ref{joint}). Therefore, the probability of positive slope
crossing of $y=\alpha$ in $\Delta t$ is
\begin{equation}\label{level1}
\sum_{0}^{\infty} \Delta{y}^{\prime}\sum_{\alpha-{y}^{\prime}\Delta
t}^{\alpha} \Delta y p(y,{y}^{\prime}).
\end{equation}

We can replace $y$ with $\alpha$ when $\Delta t\rightarrow 0$

\begin{equation}
p(y,{y}^{\prime})=p(y=\alpha,{y}^{\prime}).
\end{equation}
Since at large values of $y$ and ${y}^{\prime}$ the probability
density function approaches zero fast enough (see Fig. \ref{joint}),
consequently expression (\ref{level1}) may be written as
\cite{newland,Tabar1}

\begin{equation}
\int_{0}^{\infty} d{y}^{\prime}\int_{\alpha-{y}^{\prime}d
t}^{\alpha} d y p(y=\alpha,{y}^{\prime})
\end{equation}
in which the integrand is no longer as a function of $y$ so that the
first integral is just $\int_{\alpha-{y}^{\prime}d t}^{\alpha} d y
p(y=\alpha,{y} ^{\prime})=p(y=\alpha,{y}^{\prime}){y}^{\prime}d t $.
Finally the probability of upcrossing
($\nu_{\alpha}^{+}$) of $y=\alpha$ is equal to \cite{newland,Tabar1}

\begin{equation}\label{level}
\nu_{\alpha}^+=\int_{0}^{\infty} p(\alpha,{y}^{\prime}){y}^{\prime}d{y}^{\prime}
\end{equation}
in which the term $p(\alpha,{y}^{\prime})$ is the joint probability
density $ p(y,{y}^{\prime})$ evaluated at $y=\alpha$.

The integration of the Eq. (\ref{level}) over all levels
demonstrates another quantity, $N_{tot}^{+}$, which shows the total
number of upcrossings for the data.
\begin{equation}\label{ntot}
N_{tot}^{+}=\int_{-\infty}^{+\infty}\nu_{\alpha}^{+} d\alpha,
\end{equation}
To study the effects of correlations or memory, $N_{sh}^{+}$ is
evaluated which gives the total number of upcrossings of time series
when it is shuffled. Here, random permutation is used for shuffling
the data. The autocorrelations are destroyed by the shuffling
procedure. Hence, by comparing $N_{tot}^{+}$ of the original data
with that of computed for shuffled data set, $N_{sh}^{+}$, we can
obtain the magnitude of correlations in the time series and this
gives useful information about the time series. In order to study
the effect of distribution function in the data with non-Gaussian
distribution function, the so-called surrogate procedure can be
applied. In the surrogate method, the discrete Fourier transform
(DFT) of the observed time series data is computed and then the
phases of each complex amplitude of the DFT are replaced with
independently distributed artificial uniform $(-\pi,+\pi)$ variates
\cite{the92,Ashkenazy}. The altered DFT is then inverse Fourier
transformed to generate a surrogate time series. The correlations in
the surrogate series could be kept unchanged (for more detail see
appendix (B) reference \cite{leila}), but the probability function
changes to a Gaussian distribution
\cite{leila,sc00,the92,the93,the97,mahsa}. To obtain information
about the effect of the phase randomization procedure on the PDF,
one can also check the results of this procedure on the magnitude
and sign series \cite{Ashkenazy}. For the surrogate time series,
$N_{su}^{+}$ can be calculated which is the total number of
upcrossings of this time series.

\begin{figure}[t]
\includegraphics[width=8cm,height=7cm,angle=0]{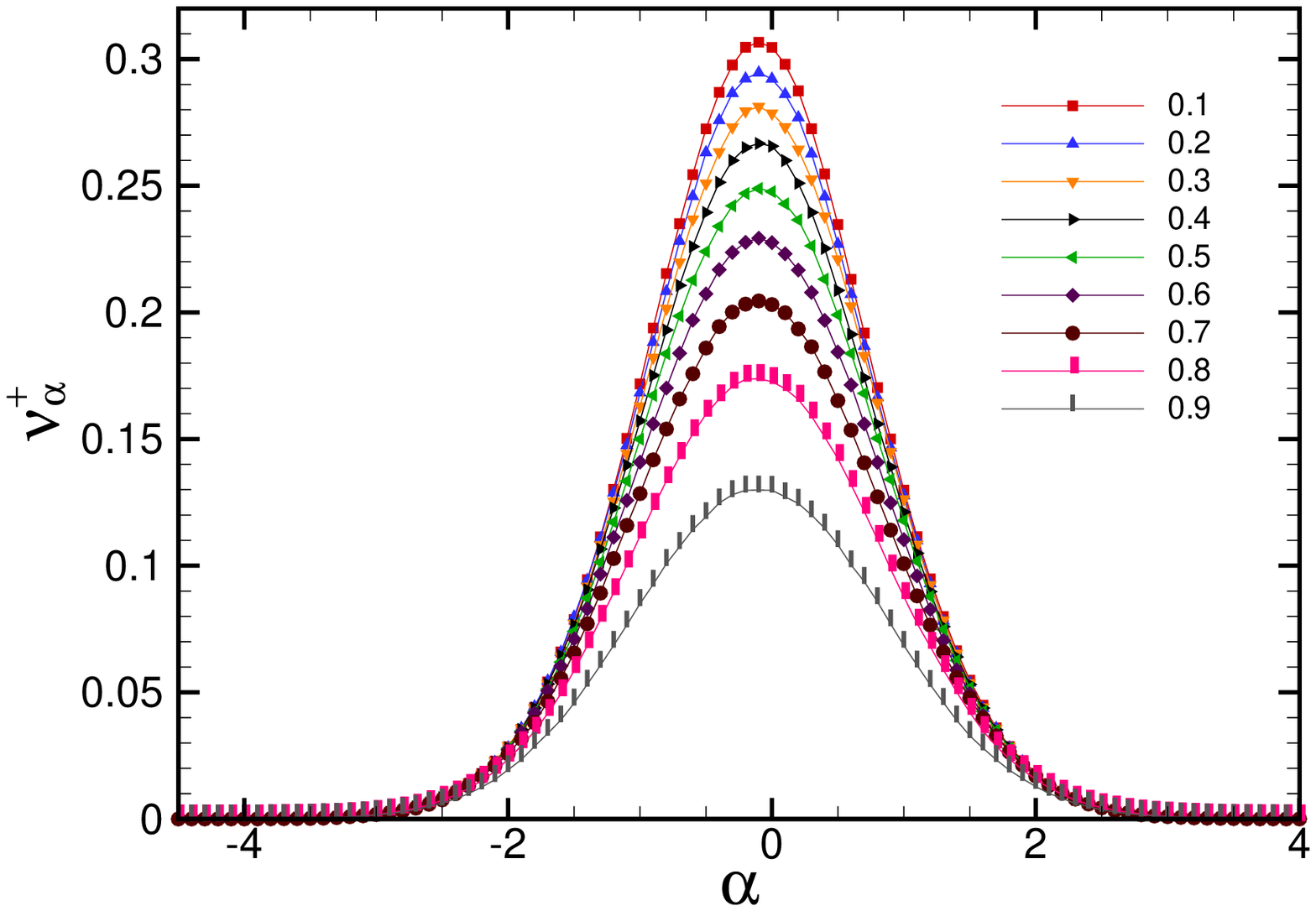}
\includegraphics[width=7.5cm,height=6.5cm,angle=0]{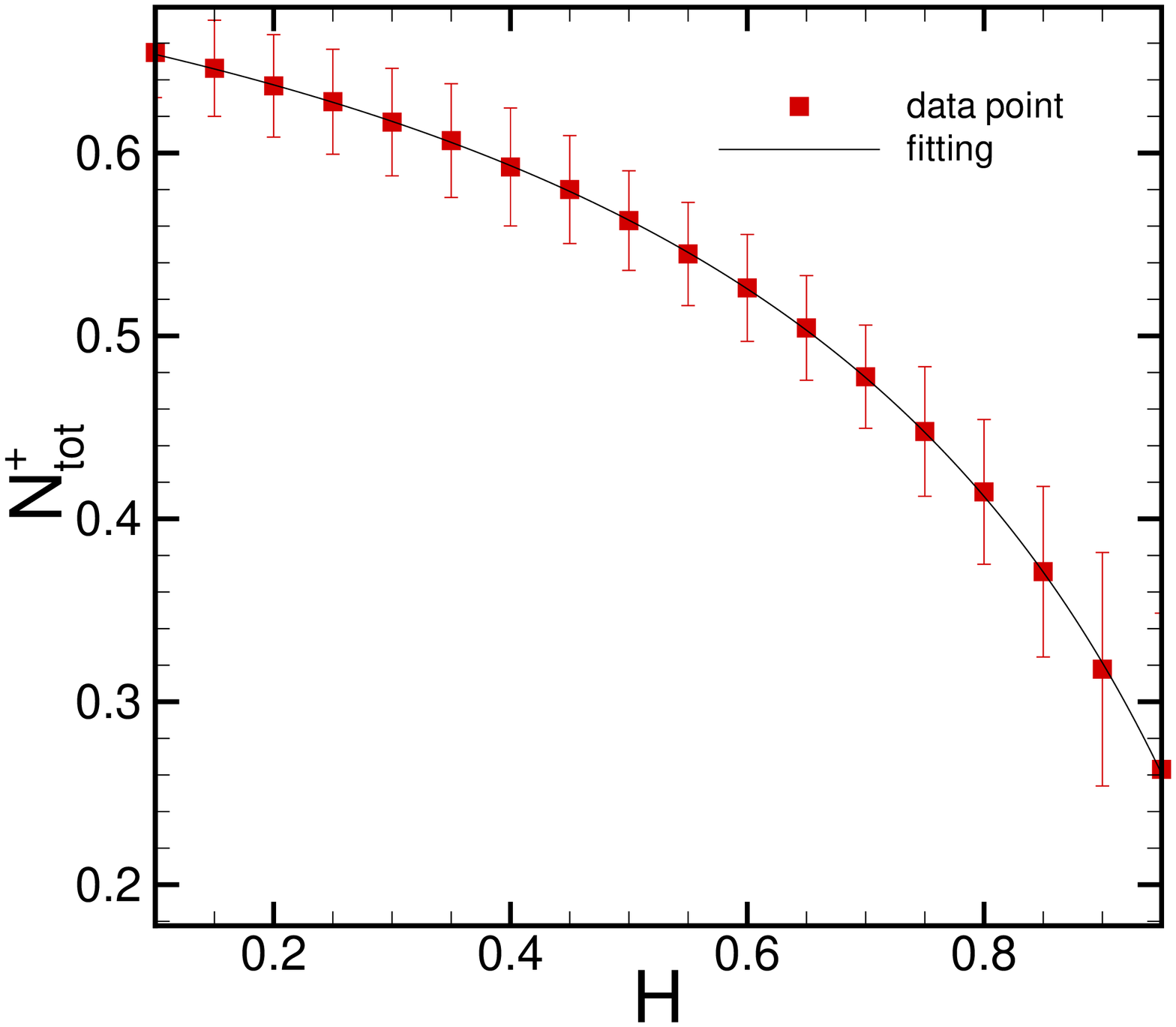}
\caption{Left panel) Positive level crossings (upcrossings) of
different fractional Gaussian noises. The average values of all fGns
have been shifted to zero and their standard deviations are one.
Right panel) $N^{+}_{tot}$ of different fGns with respect to Hurst
exponent. Symbols are given by numerical analysis and fitting
function is given by Eq. (\ref{fit1}).}\label{fig3}
\end{figure}
\begin{figure}[t]
\includegraphics[width=12cm,height=10cm,angle=0]{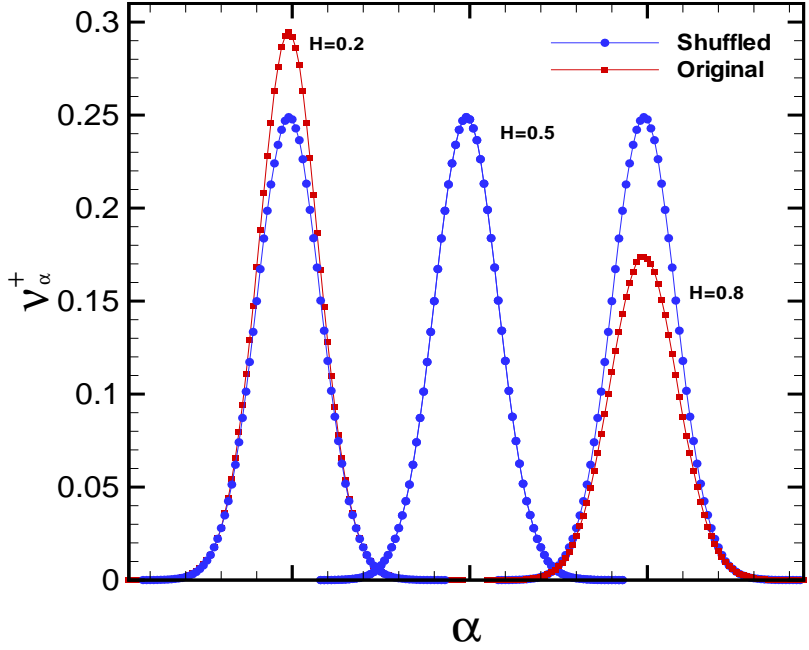}
\caption{Schematic comparison of positive level crossings for three
fGns and their shuffled.}\label{fig4}
\end{figure}

\begin{figure}[t]
\includegraphics[width=12cm,height=10cm,angle=0]{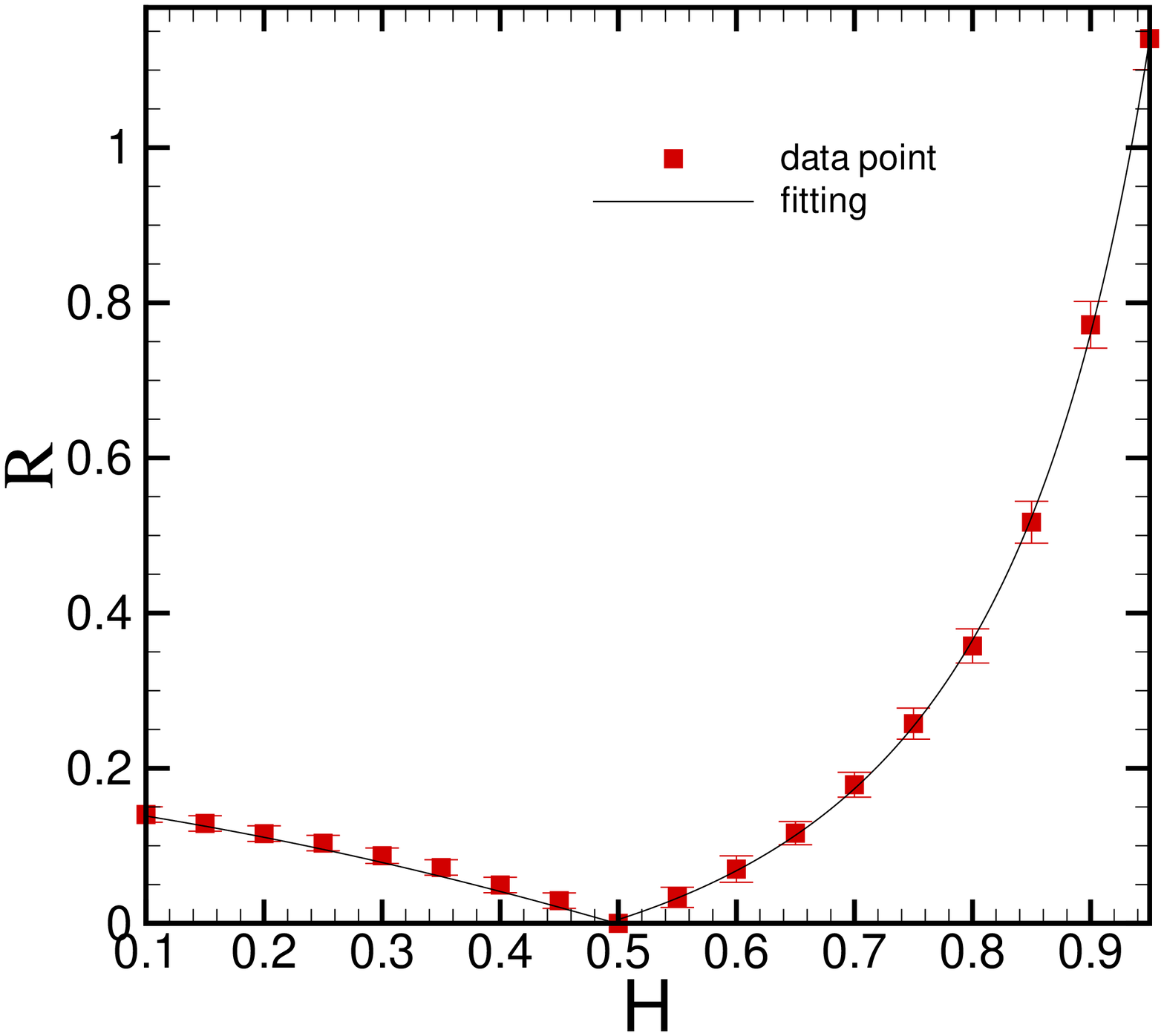}
\caption{The relative difference, $\mathcal{R}$, as a function of
$H$. Symbols are given by numerical analysis and the solid line
corresponds to the fitting function which is given by Eq.
(\ref{eq7}).}\label{fig5}
\end{figure}

\section{discussion and results}

By using the PDF and the correlation function of the data, a
stationary series can be pictured. Fractional Gaussian noises which
are generalization of ordinary discrete white Gaussian noise are
characterized by their Hurst exponent. Hurst exponent, H, gives a
quantitative measure of the long-term persistence of a signal. In
particular, the exponents $0<H<0.5$ and $0.5<H<1$ correspond to
negative (anti-correlation) and positive correlation, respectively,
while $H=0.5$ corresponds to an uncorrelated Gaussian process. For
clarifying the level crossing results of unknown empirical data,
their comparison with the results of well known fGns is emphasized.
Fractional Gaussian noises can be generated using different methods
\cite{Saupe,Havlin,Prakash,Sahimi,Pang,Makse,Mehrabi,Voss,Ausloos1}.
Here, we have used the Fourier filtering method which was fully
described in Refs. \cite{Saupe,Havlin,Prakash}. In this method, the
Fourier components of an uncorrelated sequence of random numbers are
filtered by a suitable power-law filter in order to introduce
correlations among the variables. In Fig. \ref{fig2}, we have
plotted schematically three fractional Gaussian noises which are
from up to down anticorrelated, random and correlated Gaussian
noises. For a better comparison, their plots have been shifted
vertically.

Level crossing function, $\nu^{+}_{\alpha}$, is calculated for
different fractional Gaussian noises, according to Eq. \ref{level}.
Fig. (\ref{fig3}) (left panel) shows  $\nu^{+}_{\alpha}$ as a
function of level $\alpha$ for different values of Hurst exponent.
By increasing the Hurst exponent of the fractional Gaussian noises
the total number of upcrossings decreases (see Fig. (\ref{fig3})
(right panel)). The behavior of the $N^{+}_{tot}$ elucidates the
fact that by increasing the Hurst exponent the fluctuation in the
time series decreases. The solid line shown in Fig. (\ref{fig3})
(right panel) corresponds to a rational function of the following
form that has been fitted to the curve of this figure with the
goodness of fit equals to $r=0.999$.
\begin{equation}\label{fit1}
N^+_{tot}=\frac{0.668-0.615H}{1-0.726H+0.012H^{2}}.
\end{equation}

Now it is interesting to separate the correlation and the PDF
effects by shuffling and surrogate procedures. These procedures are
used in the level crossing method to elucidate its sensitivity to
memory and PDF \cite{tabar2,mahsa}.
Since the PDF of the fractional Gaussian noise is the normal
distribution, there is no more information in the surrogate method.
By using the shuffling procedure, we can evaluate the memory of the
series. In the case of fractional Gaussian noises, the total
upcrossings (downcrossings) of the shuffled time series is increased
for the correlated time series ($H > 0.5$) while this behavior is
reversed for the anticorrelated ones (see Fig. (\ref{fig4})). For
the white noise, there is no change in the total positive level
crossings because there is no correlation in the time series. For a
better comparison, their plots have been shifted horizontally. By
comparing the difference between $N^{+}_{tot}$ and $N^{+}_{sh}$
(after shuffling), the memory of the time series can be determined.
Smaller relative difference denotes that the time series is less
correlated (anticorrelated). As it is seen (Fig. (\ref{fig4})) under
shuffling procedure the total positive level crossings
$N^{+}_{tot}$, is increased (decreased) and this confirms that the
underlying data set is correlated (anticorrelated). In order to
quantify the value of memory (correlation and anticorrelation)
embedded in the date set, we define the relative change of the total
number of upcrossings by using the $N^+_{tot}$ for original and
shuffled series as $\mathcal{R}\equiv
|N_{sh}^{+}-N_{tot}^{+}|/N_{tot}^{+}$. In Fig. (\ref{fig5}),
$\mathcal{R}$ has been plotted for various values of Hurst exponents
for different Hurst exponents. Two typical functions have been
fitted to the curve of this figure which are
\begin{eqnarray}\label{eq7}
\mathcal{R}=
\left\{   \begin{array}{cc}
                \frac{0.161-0.323H}{1-0.744H-0.441H^{2}}   & \ \ \ \ \ \ \ 0<H<0.5 \\
                \frac{-0.094+0.192H}{1-1.446H+0.500H^{2}}  & \ \ \ \ \ \ \ 0.5<H<1
\end{array}  \right.
\end{eqnarray}
with the goodness of fit $r=0.999$ for each of them.

The probability density function of stationary empirical data set
can be deviated from the Gaussian shape, consequently the total
upcrossings of the so-called surrogate data is a relevant quantity
to make our analysis complemented. In order to demonstrate, how
$\mathcal{R}$ and the other defined parameters works, we applied
this method to log-returns of three time series of financial markets
which are $S\&P 500$, Dow Jones and TSE. The results are reported in
table \ref{Tb1}. For better comparison, we have presented the
results for a white noise. Here, we have used both the shuffling and
surrogate procedures. As seen in the table, for shuffling procedure
there is not that much change in the total positive level crossing
for the two first markets which are considered as efficient markets.
But, the shuffling procedure has strong effect on TSE which is an
inefficient market. When the surrogate procedure is applied, the
total positive level crossing for $S\&P 500$ is not that much
affected but for Dow Jones and TSE this is not the case.  This is
because their probability density functions are deviated from the
Gaussian function. In order for them to be comparable to the fGns,
they should be first surrogated. In fact, correlation and PDF both
affect the level crossing method. Here, by using both of these
procedures we want to consider both their effects.
\begin{table}
\caption{\label{Tb1} Total positive level crossings of the
log-returns of three time series of financial markets ($S\&P 500$,
Dow Jones and TSE) for the original, shuffled and surrogated time
series, and the relative difference of the shuffled and the original
series one with its own pdf and the other when the pdf has been
changed to a Gaussian. For better comparison these parameters are
also presented for a white noise.}
\medskip
\begin{tabular}{|c|c|c|c|c|c|c|c|}
  \hline
   Index  &$H$ &$N^{+}_{tot}$ & $N^{+}_{sh}$   & $N^{+}_{su}$ & $|N_{sh}^{+}-N_{tot}^{+}|/N_{tot}^{+}$&$|N_{su}^{+}-N_{tot}^{+}|/N_{tot}^{+}$  \\\hline
   $S\&P500$  & 0.45   & 0.49   & 0.51  & 0.57   & 0.03 & 0.16  \\\hline
   $Dow$ $Jones$ &0.53 & 0.40   & 0.43  & 0.57   & 0.08 & 0.43  \\\hline
   $TSE$       &0.71   & 0.29   & 0.49  & 0.46   & 0.66 & 0.59  \\\hline
   $White$ $noise$&0.5 & 0.56   & 0.56  & 0.56   & 0.00 & 0.00  \\\hline
\end{tabular}
\end{table}

\section{Conclusion}

Level crossing method with no required scaling feature is a powerful
method in characterizing the time series. This method not only
introduces roughness in a new concept but also contains information
about waiting time (the time interval to observe an event again,
statistically). It can also measure the memory and non-Gaussianity.
Besides all these inferences, there was lack of a suitable measure
for better understanding the level crossing results of unknown
empirical data. Fractional Gaussian noises with normal distributions
which can be indicated by only a Hurst exponent are known samples
that could play the role of this criterion. They can fill this
vacancy by comparing their level crossing results with the results
of the unknown empirical Data. In this article, the level crossing
results of fGns are investigated. The deviation of the empirical
data results form the fGn with the same Hurst exponent is a measure
of correlation and non-Gaussianity in the empirical one.


\end{document}